\documentclass[aps,twocolumn,showpacs]{revtex4}
\usepackage{graphicx}
\usepackage{epsfig}
\usepackage{color}
\begin{document}

\title{Thermal orientation of electron spins}
\author{S.\,A.\,Tarasenko}
\affiliation{A.F.~Ioffe Physico-Technical Institute, Russian
Academy of Sciences, 194021 St.~Petersburg, Russia}





\begin{abstract}
It is shown that the spin orientation of free electrons occurs in
low-symmetry semiconductor structures if only the electron gas is
driven out of thermal equilibrium with the crystal lattice. The
proposed mechanism of such a thermal orientation of electron spins
is based on spin dependence of the electron-phonon interaction
which tends to restore equilibrium. The microscopic theory of the
effect is developed here for asymmetric (110)-grown quantum wells
where the electron gas heating leads to the spin orientation along
the $[1\bar{1}0]$ axis in the quantum well plane.
\end{abstract}

\maketitle

\section{Introduction}\label{sect1}

The spin-related phenomena in semiconductors have been attracting
considerable attention since the discovery of the optical
orientation of electron and nuclear spins and the basic mechanisms
of spin relaxation~\cite{optor}. Much effort in this field is
currently focused on the development of novel methods of spin
orientation of free carriers ranging from the spin Hall
effect~\cite{DP71b} and spin-dependent tunneling~\cite{Perel03} to
the optical orientation by linearly polarized
light~\cite{Tarasenko05}. Here we show that the spin orientation
of free electrons can be achieved in semiconductor nanostructures
by simple electron gas heating. We demonstrate that, in
nanostructures of sufficiently low space symmetry, the
electron-phonon interaction tending to restore equilibrium between
the electrons and the crystal lattice is spin-dependent and leads
to the spin orientation of carriers. Such a thermal orientation of
electron spins is considered here for asymmetric (110)-grown
quantum wells (QWs) where the electron gas heating leads to the
spin orientation along the $[1\bar{1}0]$ axis in the QW plane.

The possibility to achieve the spin polarization of carriers
$\mathbf{S}$ caused by the disturbance of thermal equilibrium in
asymmetric (110)-grown QWs follows from symmetry analysis. Indeed,
such structures belong to the point group $C_s$ that contains only
two symmetry elements, namely, identity and a mirror plane
perpendicular to the $x$ axis. Here we use the following
coordinate frame: $x \| [1\bar{1}0]$ and $y \| [00\bar{1}]$ are
the in-plane axes, and $z \| [110]$ is the growth direction.
Reflection by the mirror plane changes the sign of the $y$ and $z$
components of the spin axial vector $\mathbf{S}$ but does not
modify the $x$ component. Therefore, the spin component $S_x$ is
an invariant in asymmetric (110)-grown QWs suggesting that the
spin polarization along the $x$ axis can emerge if the electron
gas is driven out of equilibrium by any means. Particularly, in
the case of disturbance of thermal equilibrium between the
electrons and the crystal lattice, the spin orientation can
phenomenologically be described by
\begin{equation}
 S_{x} \propto \frac{\Delta T}{T_e} \:,
\end{equation}
where $\Delta T=T_e-T_0$, $T_e$ and $T_0$ are the electron and
lattice temperatures, respectively.

\section{Microscopic model}\label{sect2}

The microscopic model of the thermal orientation of electron spins
is caused by the energy relaxation of hot carriers and includes
two stages which are illustrated in Fig.~1(a) and Fig.~1(b).

In the first stage [Fig.~1(a)], the carriers lose a part of their
kinetic energy by emitting phonons. Such energy relaxation
processes shown by curved arrows are
spin-dependent~\cite{Ivchenko83}. In QW structures without an
inversion center, spin-orbit interaction adds an asymmetric term
to the probability of electron scattering by phonons which is
linear in the wave vector components~\cite{Ganichev06}. As is
shown in the next Section, the dominant spin-dependent
contribution to the probability of electron scattering in
(110)-grown QWs is proportional to $\sigma_{z}(k_{x}+k'_{x})$,
where $\sigma_{z}$ is the Pauli matrix, $k_{x}$ and $k'_{x}$ are
components of the initial and scattered electron wave vectors.
\begin{figure}[b]
\leavevmode \epsfxsize=0.95\linewidth
\centering{\epsfbox{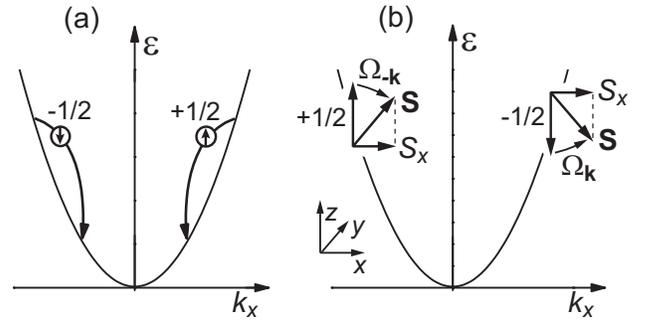}} \caption{Microscopic origin of the
electron gas spin orientation due to the energy relaxation of hot
carriers. (a) Spin-dependent asymmetry of the electron-phonon
interaction followed by (b) spin precession in the effective
magnetic field leads to the spin orientation of electrons along
the $x$ axis.}
\end{figure}
Due to the spin-dependent asymmetry of the electron-phonon
interaction, electrons with the spin "up" (along the $z$ axis)
predominantly vacate the excited states with positive $k_{x}$
while electrons with the spin "down" vacate the excited states
with negative $k_{x}$. This leads to a nonequilibrium distribution
where the spin-up hot carriers occupy mainly the left-hand branch
of the dispersion curve (carriers with the opposite spin
orientation have gone to the subband bottom) while the spin-down
carriers occupy mainly the right-hand branch.

In the second stage [Fig.~1(b)], a net spin orientation of the
electron gas appears as a result of the subsequent spin precession
of nonequilibrium carriers in the effective magnetic field induced
by the spin-orbit interaction~\cite{DP71}. The field has nonzero
in-plane components in asymmetrically growth QWs, e.g., due to the
Rashba effect~\cite{Bychkov84}. Therefore, the spins of
nonequilibrium carriers, directed along or opposite to the $z$
axis after the phonon emission, will precess in the effective
field as shown in Fig.~1(b). Note that electrons with the initial
spin $-1/2$ and wave vector $k_{x}>0$ are acted upon by the
effective field with the Larmor frequency
$\mathbf{\Omega}_{\mathbf{k}}$, while particles with the initial
spin $+1/2$ and the negative wave vector feel the field with the
Larmor frequency $\mathbf{\Omega}_{-\mathbf{k}}$. The effective
magnetic field induced by spin-orbit interaction is an odd
function of the wave vector, therefore,
$\mathbf{\Omega}_{-\mathbf{k}}=-\mathbf{\Omega}_{\mathbf{k}}$, and
the spins of particles with positive and negative values of
$k_{x}$ rotate in opposite directions. This gives rise to the spin
component $S_{x}>0$ for all hot electrons in the subband and the
net spin polarization of the electron gas.

We note that the electron spin caused by the energy relaxation
processes is nonequilibrium and, therefore, can be rotated by an
external static magnetic field, similarly to the Hanle effect in
optics.

\section{Theory}\label{sect3}

The spin-dependent asymmetry of the electron-phonon interaction
can be obtained if one takes into account both
$\mathbf{k}\cdot\mathbf{p}$ admixture of the valence-band states
to the conduction-band wave function and the phonon-induced
interband coupling. To first order in the
$\mathbf{k}\cdot\mathbf{p}$ theory, where $\mathbf{k}=(k_x,k_y)$
is the in-plane wave vector, the electron wave function in a
(110)-grown QW has the form
\begin{equation}\label{kp_func}
\Psi_{\mathbf{k}}(\mathbf{r}) = S \psi + X^\prime
\frac{v_{x}+v_{z}}{\sqrt{2}} + Y^\prime
\frac{v_{z}-v_{x}}{\sqrt{2}} - Z^\prime v_{y} \:.
\end{equation}
Here $S$ and $X^\prime$, $Y^\prime$, $Z^\prime$ are the Bloch
functions of the conduction and valence bands at the $\Gamma$
point of the Brillouin zone, $\psi$ and $\mathbf{v}=(v_x,v_y,v_z)$
are the envelope spinors. The Bloch functions $X^\prime$,
$Y^\prime$, and $Z^\prime$ in Eq.~(\ref{kp_func}) are referred to
the cubic axes $x' \|[100]$, $y' \| [010]$, and $z' \| [001]$,
respectively. The envelope spinors of the valence-band states are
related to the conduction-band spinor by
\begin{equation}\label{v_kp}
\mathbf{v} = -\frac{\hbar P_{cv}}{3m_0}
\,\frac{(3E_g+2\Delta_{so}) \mathbf{k} + i \Delta_{so}
[\mathbf{\sigma}\times\mathbf{k}]}{E_g(E_g+\Delta_{so})} \, \psi
\:,
\end{equation}
where $P_{cv}=\langle S| p_{z'}|Z^\prime \rangle$ is the interband
matrix element of the momentum operator, $m_0$ is the free
electron mass, $E_g$ is the band gap, $\Delta_{so}$ is the
valence-band spin-orbit splitting, and $\mathbf{\sigma}$ is the
vector of the Pauli matrices.

We consider the electron scattering by acoustic phonons due to the
deformation-potential mechanism. In cubic noncentrosymmetric
crystals such as zinc-blende-type semiconductors, the strain
induces a coupling between the conduction-band and valence-band
states~\cite{optor,Ivchenko04}. The matrix elements of such a
coupling have the form $V_{S,X^\prime}=\Xi_{cv} u_{y'z'}$,
$V_{S,Y^\prime}=\Xi_{cv} u_{x'z'}$, $V_{S,Z^\prime}=\Xi_{cv}
u_{x'y'}$, where $\Xi_{cv}$ is the interband constant of the
deformation potential, and $u_{\alpha\beta}$ are the strain tensor
components used here in the primed coordinate system. It is the
strain-induced interband coupling together with the spin-orbit
splitting of the valence band that leads to spin-dependent
asymmetry of the electron-phonon interaction. Taking into account
$\mathbf{k}\cdot\mathbf{p}$ mixing given by Eqs.~(\ref{kp_func})
and (\ref{v_kp}) and allowing for the interband coupling, we
derive for the Hamiltonian of the electron-phonon interaction in
(110)-grown quantum wells
\begin{equation}\label{V_elphon}
V_{\mathrm{el-phon}}(\mathbf{k}',\mathbf{k}) = \Xi_c \sum_{\alpha}
u_{\alpha\alpha} + \xi \, \Xi_{cv} \left\{ (k_{x} + k'_{x}) \times
\right.
\end{equation}
\vspace{-0.8cm}
\[
\left. [\sigma_{z} (u_{zz}-u_{xx})/2 - \sigma_{y} u_{yz}] + (k_{y}
+ k'_{y}) [\sigma_{x} u_{yz} + \sigma_{z} u_{xy}] \right\} \:,
\]
where $\Xi_c$ is the intraband constant of the deformation
potential responsible for the dominant (spin-independent) part of
the electron-phonon interaction, and $\xi$ is the coefficient
given by
\begin{equation}\label{xi}
\xi = \frac{i\hbar P_{cv}}{3m_0}
\frac{\Delta_{so}}{E_g(E_g+\Delta_{so})} \:.
\end{equation}
In deriving Eq.~(\ref{V_elphon}), we have expressed the strain
tensor components in the primed axes via those in the QW
coordinate system $xyz$ by using the equalities
$u_{x'y'}=(u_{zz}-u_{xx})/2$,
$u_{x'z'}=-(u_{xy}+u_{yz})/\sqrt{2}$,
$u_{y'z'}=(u_{xy}-u_{yz})/\sqrt{2}$.

The probability of electron scattering is determined by squared
matrix elements of the electron-phonon
interaction~(\ref{V_elphon}). The dominant contribution to
spin-dependent asymmetry of the electron scattering in quantum
wells is given by terms proportional to the $u_{zz}$ component of
the phonon-induced strain tensor. This is because the strain
tensor components depend on the wave vector and polarization of
the phonon involved, and the in-plane component of the phonon wave
vector $q_{\parallel}=|\mathbf{k}-\mathbf{k}'|$ is typically much
smaller than the out-of-plane component $q_z \sim \pi/a$, where
$a$ is the QW width. Thus, the principle contribution to the
scattering asymmetry in (110)-grown QW structures is proportional
to $\sigma_z (k_x+k'_x)$. This term is taken into account in
calculations which follow. We note that $\mathbf{k}$-linear terms
in the matrix elements of the electron-phonon interaction can also
be obtained in the second order in the $\mathbf{k}\cdot\mathbf{p}$
theory, as was done in Ref.~\cite{Averkiev02} for the electron
scattering by charge impurities. However, these terms lead to no
essential contribution to asymmetry of the electron scattering by
phonons nor the spin orientation caused by the energy relaxation
processes.

As is shown in Section~\ref{sect2}, spin-dependent asymmetry of
the scattering processes followed by the precession of electron
spins in the effective magnetic field leads to spin orientation of
the electron gas. We assume that the spin relaxation time of
carriers is much longer than the thermalization time controlled by
electron-electron collisions which is in turn much longer than the
momentum relaxation time $\tau_p$, and $\Omega_{\mathbf{k}} \tau_p
\ll 1$. Then, the spin generation rate is given
by~\cite{Tarasenko05}
\begin{equation}\label{S_dot}
\dot{\mathbf{S}} = \sum_{\mathbf{k}} \tau_p \,
[\mathbf{\Omega}_{\mathbf{k}}\times\mathbf{g}_{\mathbf{k}}] \:,
\end{equation}
where $\mathbf{g}_{\mathbf{k}}=\mathrm{Tr}[\mathbf{\sigma}
G(\mathbf{k})]/2$, $G(\mathbf{k})$ is the spin matrix describing
the carrier redistribution in $\mathbf{k}$-space due to the
scattering by phonons. Components of the matrix $G(\mathbf{k})$
have the form~(see, e.g., Ref.~\cite{Ivchenko04})
\begin{equation}\label{generation}
G_{ss'}(\mathbf{k}) = \frac{2\pi}{\hbar} \sum_{s1,\mathbf{k}1}
\sum_{\mathbf{q},\pm} \times
\end{equation}
\vspace{-0.5cm}
\[
\left\{ V_{s\mathbf{k},s1\mathbf{k}1}^{(\pm)}
V_{s'\mathbf{k},s1\mathbf{k}1}^{(\pm)*} \,
f({\varepsilon_{\mathbf{k}1}}) [1 - f(\varepsilon_{\mathbf{k}})]
\,\delta(\varepsilon_{\mathbf{k}}-\varepsilon_{\mathbf{k}1} \pm
\hbar\omega_{\mathbf{q}}) \right.
\]
\vspace{-0.5cm}
\[
- \left. V_{s1\mathbf{k}1,s\mathbf{k}}^{(\pm)*}
V_{s1\mathbf{k}1,s'\mathbf{k}}^{(\pm)} \,
f(\varepsilon_{\mathbf{k}}) [1 - f(\varepsilon_{\mathbf{k}1}) ] \,
\delta(\varepsilon_{\mathbf{k}1}-\varepsilon_{\mathbf{k}} \pm
\hbar\omega_{\mathbf{q}}) \right\} \:,
\]
where $V_{s\mathbf{k},s1\mathbf{k}1}^{(\pm)}$ is the matrix
element of the electron scattering assisted by emission ($+$) or
absorption ($-$) of a phonon, $f(\varepsilon_{\mathbf{k}})$ is the
distribution function of carriers,
$\varepsilon_{\mathbf{k}}=\hbar^2k^2/(2m^*)$ is the kinetic
energy, and $\omega_{\mathbf{q}}$ is the phonon frequency. Note
that spin-orbit splitting of the energy spectrum is neglected in
Eq.~(\ref{generation}).

In (110)-grown QW structures, components of the Larmor frequency
of the effective magnetic field have the form
\begin{equation}
\mathbf{\Omega}_{\mathbf{k}} = \frac{2}{\hbar}
(\gamma_{xy}k_y,\gamma_{yx}k_x,\gamma_{zx}k_x) \:.
\end{equation}
The parameter $\gamma_{zx}$ is caused here by the lack of an
inversion center in the host crystal, while $\gamma_{xy}$ and
$\gamma_{yx}$ are non-zero due to the QW asymmetry only. We assume
that electrons obey the Boltzmann statistics and the lattice
temperature is not very low, $k_B T_0 \gg
\hbar\omega_{\mathbf{q}}$. Then, one derives
\begin{equation}\label{S_dot2}
\dot{S}_{x} = - \tau_p \gamma_{yx} \frac{\Xi_c \Xi_{cv}}{2\rho}
\frac{\xi m^{*2}}{\hbar^4} \frac{\Delta T}{T_e} N_e \times
\int\limits_{-\infty}^{+\infty} \left[\frac{d \, \varphi^2(z)}{d
z} \right]^2 dz \:,
\end{equation}
where $\rho$ is the crystal density,
$N_e=2\sum_{\mathbf{k}}f(\varepsilon_{\mathbf{k}})$ is the
two-dimensional electron concentration, and $\varphi(z)$ is the
function of size quantization.

In the steady state regime, when the electron temperature $T_e$
and the lattice temperature $T_0$ are held constant, the spin
density $S_x$ is determined by balance between the spin generation
and relaxation processes; $S_x=\dot{S}_{x}T_{x}$, where $T_{x}$ is
the spin relaxation time. In (001)-grown QWs, the time $T_x$ is
given by~\cite{Dyakonov86}
\begin{equation}\label{T_spin}
T_x^{-1} = - \int_{0}^{\infty} \frac{\tau_p}{f(0)} \frac{d
f(\varepsilon_{\mathbf{k}})}{d\varepsilon_{\mathbf{k}}}  \left(
\langle \Omega_{\mathbf{k},y}^2 \rangle + \langle
\Omega_{\mathbf{k},z}^2 \rangle \right) d\varepsilon_{\mathbf{k}}
\:,
\end{equation}
where the angle brackets mean averaging over directions of the
wave vector. For the Boltzmann distribution, Eq.~(\ref{T_spin})
assumes the form
\begin{equation}\label{T_spin2}
T_x^{-1} = \frac{4 m^* \tau_p}{\hbar^4} (\gamma_{yx}^2 +
\gamma_{zx}^2) k_B T_e \:.
\end{equation}
Finally, we obtain for the steady spin density
\begin{equation}\label{S_x}
S_{x} = - \frac{m^* \xi}{8 \rho} \frac{\gamma_{yx}\, \Xi_c
\Xi_{cv}}{\gamma_{yx}^2+\gamma_{zx}^2} \frac{\Delta T N_e}{k_B
T_e^2} \times \int\limits_{-\infty}^{+\infty} \left[\frac{d \,
\varphi^2(z)}{d z} \right]^2 dz \:.
\end{equation}
The estimation for the average electron spin gives
$S_x/N_e\sim10^{-5}$ for the electron temperature $T_e=100$~K, the
ratio $\Delta T/T_e \approx 1$, the quantum well width
$a=100$~\AA, and band parameters $m^*=0.07\,m_0$, $\xi=0.4$~\AA,
$\gamma/\hbar=10^5$~cm/s, $\Xi_c=-8$~eV,
$\Xi_{cv}=3$~eV~\cite{optor} corresponding to GaAs-based QW
structures. Thus, for the carrier density $N_e=10^{11}$~cm$^{-2}$,
the spin density $S_x$ is of the order of $10^6$~cm$^{-2}$ which
is well above the experimental resolution. We also note that the
modest estimated value of the spin polarization is due to the fact
that the energy relaxation by acoustic phonons is ineffective. The
spin polarization can considerably increase if optical phonons are
involved in the energy relaxation processes.

\paragraph*{Acknowledgments.} This work was supported by the RFFI,
programs of the RAS, and the President Grant for young scientists.

\end{document}